# Shape reconstruction in X-ray tomography from a small number of projections using deformable models

Ali Mohammad–Djafari[1]    and    Ken Sauer[2]

[1] Laboratoire des Signaux et Systèmes (CNRS–SUPELEC–UPS)
École Supérieure d'Électricité
Plateau de Moulon, 91192 Gif–sur–Yvette Cedex, France

[2] Department of Electrical Engineering
University of Notre Dame
Notre Dame, IN 46556, USA

**Abstract.** X-ray tomographic image reconstruction consists of determining an object function from its projections. In many applications such as non-destructive testing, we look for a fault region (air) in a homogeneous, known background (metal). The image reconstruction problem then becomes the determination of the shape of the default region. Two approaches can be used: modeling the image as a binary Markov random field and estimating the pixels of the image, or modeling the shape of the fault and estimating it directly from the projections. In this work we model the fault shape by a deformable polygonal disc or a deformable polyhedral volume and propose a new method for directly estimating the coordinates of its vertices from a very limited number of its projections. The basic idea is not new, but in other competing methods, in general, the fault shape is modeled by a small number of parameters (polygonal shapes with very small number of vertices, snakes and deformable templates) and these parameters are estimated either by least squares or by maximum likelihood methods. We propose modeling the shape of the fault region by a polygon with a large number of vertices, allowing modeling of nearly any shape and estimation of its vertices' coordinates directly from the projections by defining the solution as the minimizer of an appropriate regularized criterion. This formulation can also be interpreted as a maximum a posteriori (MAP) estimate in a Bayesian estimation framework. To optimize this criterion we use either a simulated annealing or a special purpose deterministic algorithm based on iterated conditional modes (ICM). The simulated results are very encouraging, especially when the number and the angles of projections are very limited.

## 1. Introduction

Tomographic image reconstruction in non destructive testing (NDT) is recent and consists in determining an object $f(x,y)$ from its projects $p(r,\phi)$:

$$p(r,\phi) = \iint f(x,y)\delta(r - x\cos\phi - y\sin\phi)\,\mathrm{d}x\,\mathrm{d}y \qquad (1)$$

In many image reconstruction applications, especially in NDT, we know that $f(x, y)$ has a constant value $c_1$ inside a region $P$ (fault region) and another constant value $c_2$ outside that region (safe or background region), e.g. metal & air.

$$f(x, y) = \begin{cases} c_1 & \text{if } (x, y) \in P, \\ c_2 & \text{elsewhere} \end{cases} \quad (2)$$

The image reconstruction problem becomes then the determination of the shape of the fault region $P$. In this work, without loss of generality, we assume that $c_1 = 1$ and $c_2 = 0$ and model the shape of the object by its contour.

There has been many works in image reconstruction dealing with this problem. To emphasis the originality and the place of this work, we give here a summary of the different approaches for this problem:

- The first approach consists in discretizing the equation (1) to obtain:

$$\boldsymbol{p} = \boldsymbol{H}\boldsymbol{f} + \boldsymbol{n} \quad (3)$$

where, $\boldsymbol{f}$ is the discretized values of the object $f(x, y)$ (the pixel values of the image), $\boldsymbol{p}$ is values of the projection data $p(r, \phi)$, $\boldsymbol{n}$ is a vector to represent the modeling and measurement errors (noise) and $\boldsymbol{H}$ the discretized Radon operator. Then the solution is defined as the minimizer of a compound criterion

$$J(\boldsymbol{f}) = Q(\boldsymbol{p} - \boldsymbol{H}\boldsymbol{f}) + \lambda \Omega(\boldsymbol{f}), \quad (4)$$

where $\lambda$ is the regularization parameter and $Q$ and $\Omega$ has to be chosen appropriately to reflect our prior knowledge on the noise and on the image. This is the classical regularization approach of general image reconstruction problem. One can also interpret $J(\boldsymbol{f})$ as the maximum a posteriori (MAP) criterion in the Bayesian estimation framework where $\exp\left[-Q(\boldsymbol{f})\right]$ represents the likelihood term and $\exp\left[-\Omega(\boldsymbol{f})\right]$ the prior probability law.

This approach has been used with success in many applications (e.g. [1, 2, 3, 4]) but the cost of its calculation is huge due to the great dimension of $\boldsymbol{f}$. Many works have been done on choosing appropriate regularization functionals or equivalently appropriate prior probability laws for $\boldsymbol{f}$ to enforce some special properties of the image such as smoothness, positivity or piecewise smoothness [5, 6, 7, 8, 9]. Among these, one can mention mainly two types of functions for $\Omega(\boldsymbol{f})$:

Entropic laws:

$$\Omega(\boldsymbol{f}) = \sum_{j=1}^{N} \phi(f_j) \quad \text{with} \quad \phi(x) = \{|x|^p, -x\log x, \log x, \cdots\} \quad (5)$$

Markovian laws:

$$\Omega(\boldsymbol{f}) = \sum_{j=1}^{N} \sum_{i \in \mathcal{N}_j} \phi(f_j, f_i) \quad (6)$$

with convex potential functions:

$$\phi(x,y) = \left\{ |x-y|^p, -|x-y|\log\frac{x}{y}, \log\cosh|x-y|, \cdots \right\} \tag{7}$$

or non convex potential functions:

$$\phi(x,y) = \left\{ \min\{|x-y|^2, 1\}, \frac{-1}{1+|x-y|^2}, \cdots \right\} \tag{8}$$

See for example [6] for the entropic laws, [7, 5] for scale invariant markovian laws with convex potential functions and [8, 10, 9, 11] for markovian laws with non convex potential functions and other specific choices.

• The second approach consists in modeling directly the closed contour of the object as the zero-crossing of a smooth function $u(x,y)$:

$$\partial D = \{(x,y) : u(x,y) = 0\} \quad \text{and} \quad f(x,y) = \begin{cases} c_1 & \text{if } u(x,y) > 0, \\ c_2 & \text{if } u(x,y) < 0 \end{cases}, \tag{9}$$

and in defining a time evolution for $u$ (consequently the corresponding contour $\partial D$ and function $f$) such that

$$\partial D(t) = \{(x,y) : u(x,y,t) = 0\} \tag{10}$$

and such that as time $t \mapsto \infty$ we arrive at a function $u(x,y)$ such that the associated $f(x,y)$ is a solution to the inverse problem in a least square (LS) sense. This means that the evolution of $u$ and consequently the corresponding contour $\partial D$ and object $f$ is such that the LS criterion

$$J(f) = \|p - R(f)\|^2 \tag{11}$$

decreases during the evolution. In this approach the function $u$ is assimilated to a surface (of heat front wave) and the evolution of the contours $\partial D$ is constraint to be perpendicular to the surface. This means that the variation of the interior region $(\delta x, \delta y)$ is such that

$$(\delta x, \delta y) = \alpha(x,y,t)\frac{\nabla u}{|\nabla u|} \tag{12}$$

This is the Level-Set approach originally developed by Osher and Sethian [12] for problems involving the motion of curves and surfaces and then adapted, used and referred as *snakes* or *active contour models* by many authors in computer vision [13, 14] and recently for inverse problems [15]. See also [16, 17] for the application of the deformable templates in tomographic image reconstruction.

This approach also needs pixel or voxel representation of the image and its calculation cost is huge, specially in 3D imaging systems. We are presently working on this approach trying to extend it for minimizing a regularized criterion in place of the the LS criterion (11) and implementing it in 2D and 3D tomographic image reconstruction.

• The third approach starts by giving a parametric model for the object and then tries to estimate these parameters using least squares (LS) or maximum likelihood (ML) methods. In general, in this approach one chooses a parametric model such as superposition of circular or elliptical homogeneous regions to be able to relate analytically the projections to these parameters. For example, for a superposition of elliptical homogeneous regions we have:

$$f(x,y) = \sum_{k=1}^{K} d_k f_k(x - \alpha_k, y - \beta_k) \tag{13}$$

$$\text{with } f_k(x,y) = \begin{cases} 1 & \text{if } (x/a_k)^2 + (y/b_k)^2 < 1, \\ 0 & \text{elsewhere} \end{cases} \tag{14}$$

where $\boldsymbol{\theta} = \{d_k, \alpha_k, \beta_k, a_k, b_k, k = 1, \cdots, K\}$ is a vector of parameters defining the parametric model of the image (density values, coordinates of the centers and the two diameters of the ellipses). It is then easy to calculate analytically the projections and the relation between the data and the unknown parameters becomes:

$$p(r, \phi) = h(r, \phi; \boldsymbol{\theta}) + n(r, \phi) \tag{15}$$

where $h(r, \phi; \boldsymbol{\theta})$ has an analytic expression in $\boldsymbol{\theta}$. The LS or the ML estimate when the noise is assumed to be zero mean, white and Gaussian, is then given by:

$$\widehat{\boldsymbol{\theta}} = \arg\min_{\boldsymbol{\theta}} \left\{ \|p(r, \phi) - h(r, \phi; \boldsymbol{\theta})\|^2 \right\} \tag{16}$$

This approach has also been used with success in image reconstruction [18, 19, 20, 21]. But, the range of applicability of these methods is limited to the cases where the parametric models are actually appropriate.

• The fourth approach which is more appropriate to our problem of shape reconstruction, consists in modeling directly the contour of the object by a function, say $g(\theta)$ in a cylindrical coordinates $(\rho, \theta)$ such as:

$$D = \left\{ (x, y) : x^2 + y^2 = g^2(\theta) \right\}. \tag{17}$$

The next step is then to relate the projections $p(r, \phi)$ to $g(\theta)$ which, in this case is:

$$p(r, \phi) = \int_0^{2\pi} \int_0^{g(\theta)} \delta(r - \rho \cos(\phi - \theta)) \rho \, d\rho \, d\theta. \tag{18}$$

and finally to discretize this relation to obtain:

$$\boldsymbol{p} = \boldsymbol{h}(\boldsymbol{g}) + \boldsymbol{n} \tag{19}$$

where $\boldsymbol{g}$ represents the discretized values of $g(\theta)$ defining the contour of the object and $\boldsymbol{h}(\boldsymbol{g})$ represents the discretized version of the nonlinear operator (18) relating projection data $\boldsymbol{p}$ and $\boldsymbol{g}$. Then, one defines the solution as the argument which minimizes

$$J(\boldsymbol{g}) = ||\boldsymbol{p} - \boldsymbol{h}(\boldsymbol{g})||^2 + \lambda\Omega(\boldsymbol{g}), \quad (20)$$

where $\Omega(\boldsymbol{g})$ has to be chosen appropriately to reflect some regularity property of the object's contour.

In this case also one can consider $J(\boldsymbol{g})$ as the MAP criterion with $Q(\boldsymbol{g}) = ||\boldsymbol{p} - \boldsymbol{h}(\boldsymbol{g})||^2$ as the likelihood term and $\Omega(\boldsymbol{g})$ as the prior one.
This approach has been used in image restoration [22], but it seems to be new in image reconstruction applications and the proposed method in this work is in this category. The originality of our work is to model the contour of the object by a piecewise linear function which means that the object is modeled as a polygonal region whose vertices are estimated directly from the projection data.

## 2. Proposed method

In this paper we propose to model the contour of the object (fault region) as a periodic piecewise linear function or equivalently to model the shape of the object as a polygonal region with a great number $N$ of vertices to be able to approximate any shape. Then we propose to estimate directly the coordinates $\{(x_j, y_j), j = 1, \cdots, N\}$ of the vertices of this polygonal region from the projection data.

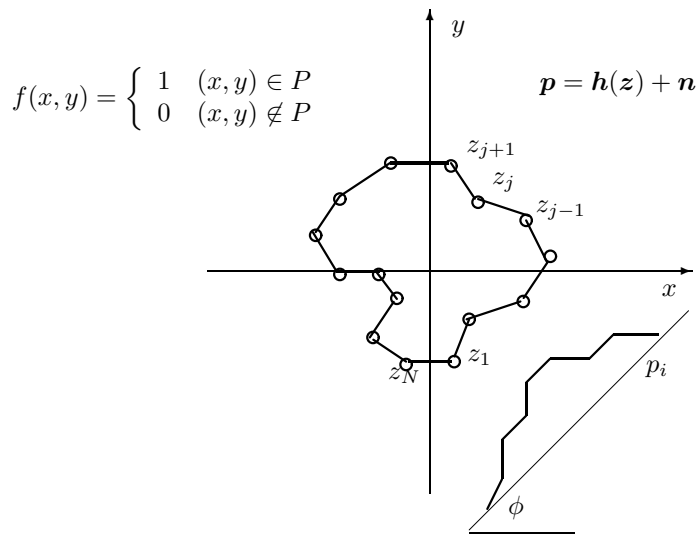

Figure 1: Proposed shape reconstruction modeling.

The idea of modeling the shape of the object as a polygonal region is not new and some works have been done in image reconstruction applications. See for example Milanfar, Karl & Willsky [23, 24, 25], but in general, in these works, either a hypothesis of convexity of the polygonal region has been used which is very restrictive in real applications or the number of vertices of the polygon is restricted to a very small number. In our work we do not make neither of these hypothesis; the polygon can be convex or not and we choose $N$ sufficiently great to to be able to approximate appropriately any shape.

The solution is then defined as the minimizer the following criterion

$$J(\boldsymbol{z}) = ||\boldsymbol{p} - \boldsymbol{h}(\boldsymbol{z})||^2 + \lambda \Omega(\boldsymbol{z}), \tag{21}$$

where $\boldsymbol{z} = \boldsymbol{x} + i\boldsymbol{y}$ is a complex vector whose real and imaginary parts represent the $x$ and the $y$ coordinates of the polygon vertices, $\boldsymbol{h}(\boldsymbol{z})$ represents the direct operator which calculates the projections for any given $\boldsymbol{z}$ and $\Omega(\boldsymbol{z})$ is chosen to be a function which reflects the regularity of the object contour. One can for example choose the following function:

$$\Omega(\boldsymbol{z}) = \left( \sum_{j=1}^{N} |z_{j+1} - z_j| - 2\pi R_0 \right)^2 \tag{22}$$

which favors a shape whose contour length is near a prior known value $2\pi R_0$. In this work we used the following function:

$$\Omega(\boldsymbol{z}) = \sum_{j=1}^{N} |z_{j-1} - 2z_j + z_{j+1}|^2 = 4 \sum_{j=1}^{N} |z_j - (z_{j-1} - z_{j+1})/2|^2, \tag{23}$$

which favors a shape whose local curvature is limited. Note that $|z_j - (z_{j-1} - z_{j+1})/2|$ is just the Euclidian distance between the point $z_j$ and the midpoint of the line segment passing through $z_{j-1}$ and $z_{j+1}$ and so this choice favors a shape whose local curvature is limited. We can also give a probabilistic interpretation to this choice. In fact we can consider $z_j$ as random variables with the following Markovian law:

$$p(z_j|\boldsymbol{z}) = p(z_j|z_{j-1}, z_{j+1}) \propto \exp\left[ -\frac{1}{2\sigma^2} |z_j - (z_{j-1} - z_{j+1})/2|^2 \right]. \tag{24}$$

Other functions are possible and are studied in this work.

In both cases, the criterion $J(\boldsymbol{z})$ is multi-modal essentially due to the fact that $\boldsymbol{h}(\boldsymbol{z})$ is a nonlinear function of $\boldsymbol{z}$. Calculating the optimal solution corresponding to the global minimum of (21) needs then carefully designed algorithms. For this we propose the two following strategies:

• The first is to use a global optimization technique such as simulated annealing. This technique has given satisfactory results as it can be seen from the simulations in the next section. However, this algorithm needs a great number of iterations and some skills for choosing the first temperature and cooling schedule, but the overall calculations is not very important due to the fact that, in this algorithm, at each iteration only one of the vertices $z_j$ is changed. So, at each step we need to calculate the variation of the criterion due to this change which can be done with a reasonable cost.

• The second is to find an initial solution in the attractive region of the global optimum and to use a local descent type algorithm such as the ICM (Iterated conditional modes) of Besag [26, 27, 28] to find the solution.

The main problem here is how to find this initial solution. For this, we used a moment based method proposed by Milanfar, Karl & Willsky [23, 24, 25] which is accurate enough to obtain an initial solution which is not very far from the optimum. The basic idea of this method is to relate the moments of the projections to the moments of a class of polygonal regions obtained by an affine transformation of a regular polygonal region, and so to estimate a polygonal region whose corners are on an ellipse and whose moments up to the second order matches those of the projections.

However, there is no theoretical proof that this initial solution will be in the attractive region of the global optimum. The next section will show some results comparing the performances of these two methods as well as a comparison with some other classical methods.

## 3. Simulation results

To measure the performances of the proposed method and keeping the objective of using this method for NDT applications where the number of projections are very limited, we simulated two cases where the objects have polygonal shapes with $N = 40$ corners (hand-made) and calculated their projections for only 5 directions

$$\phi = \{-45, -22.5, 0, 22.5, 45\} \quad \text{degrees.}$$

Then, we added some noise (white, Gaussian and centered) on them to simulate the measurement errors. The signal to noise ratio (SNR) was chosen 20dB. We define SNR as follows:

$$SNR = 10 \log \frac{\sum_i (p_i - \bar{p})^2}{\sum_i (n_i - \bar{n})^2}$$

where $p_i$ and $n_i$ are respectively data and noise samples and $\bar{p}$ and $\bar{n}$ their respective mean values.

Finally, from these data we estimated the solutions by the proposed method using either the simulated annealing (SA) or the iterated conditional modes (ICM) algorithms.

Figure 2 shows these two objects and their relative simulated projections data.

In Figures 3 and 4, we give the reconstruction results obtained by the proposed method using either the SA algorithm (Figure 3) or the ICM algorithm (Figure 4). In these figures we show the original objects, the initial solutions, the intermediate solutions during the iterations and the final reconstructed objects obtained after 200 iterations.

Note that, the SA algorithm is theoretically independent of initialization while the ICM is not. However, in these figures, for the purpose of the comparison, we show the results obtained by the two algorithms with the same initialization.

To show that the method is not very sensible on the prior knowledge of the vertices number, we give in Figure 5, the reconstruction results of the object 2 in 4 cases: $N = 10$, $N = 20$ and $N = 30$ and $N = 40$. As we can remark all the reconstructed results seem satisfactory.

In Figure 6 we show a comparison between the results obtained by the proposed method and those obtained either by a classical backprojection method or by some other methods in the first approach using (3) and (4) with different regularization functionals $\Omega(\boldsymbol{f})$, more specifically:

- *Gaussian Markov models* (5) with the potential function $\phi(x,y) = |x-y|^2$ which can also be considered as a quadratic regularization method; and
- *Compound Markov models* with non convex potential functions $\phi(x,y) = \min\{|x-y|^2, 1\}$ which is a truncated quadratic potential function.

In the first case the criterion to optimize is convex and we used a conjugate gradient (CG) algorithm to find the optimized solution. In the second case the criterion is not convex and we used a Graduated non convexity (GNC) based optimization algorithm developed in [10, 9, 11] to find the solution.

Note that, these results are given here to show the relative performances of these methods in a very difficult situation where we have only five projections. In fact, in more comfortable situations (more projections uniformly distributed around the object and high SNR) all these methods, even the simplest one such as the classical backprojection will give similar and satisfactory results. Here, we compare the results obtained from the same restricted set of data.

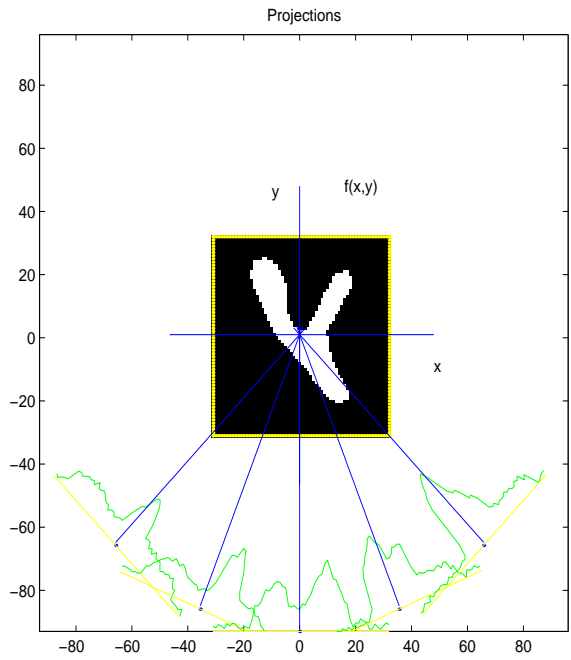
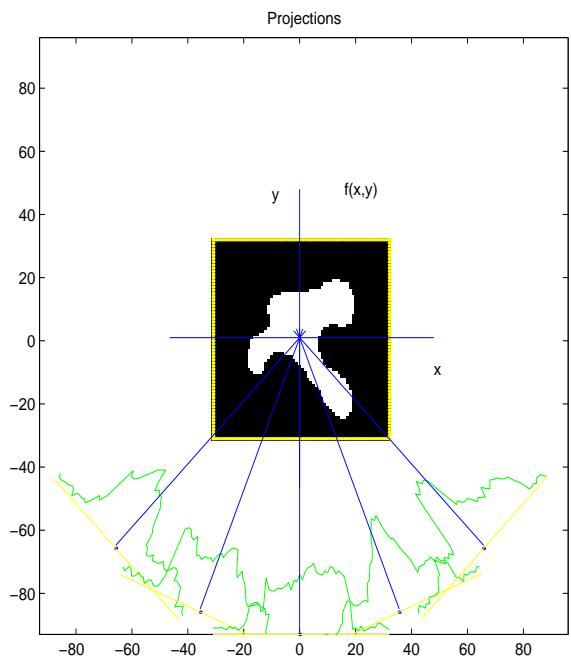

Figure 2: Original images and simulated projections data.

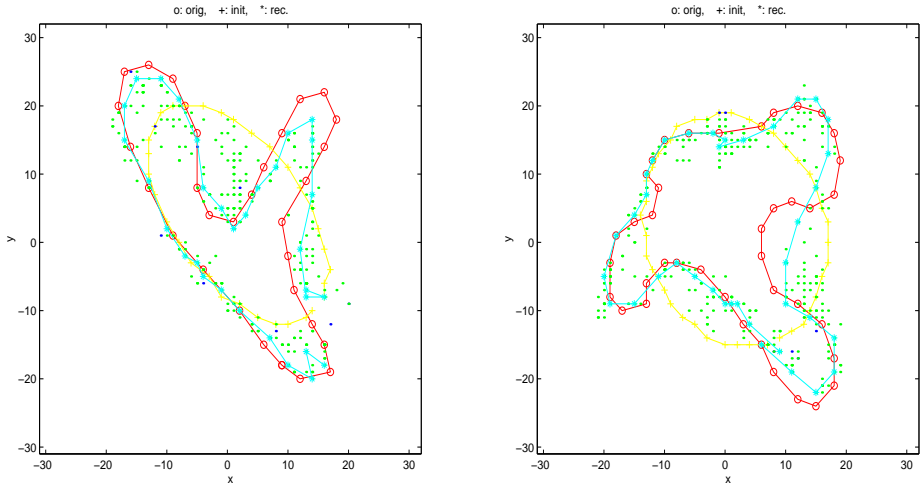

Figure 3: Reconstruction results using simulated annealing.
o) Original objects,   +) Initializations,   .) Evolution of the solutions during the iterations and
⋆) Final reconstructed objects.

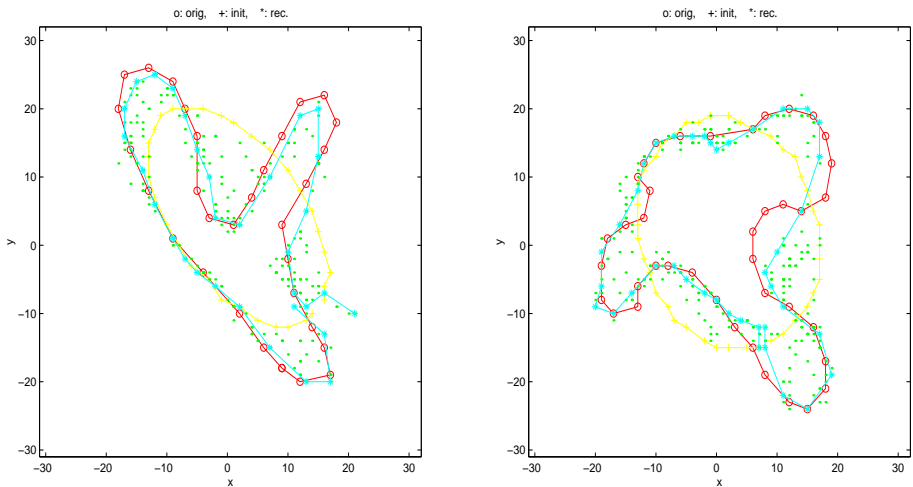

Figure 4: Reconstruction results using a moment-based initialization and a local descent (ICM) minimizer.
o) Original objects,   +) Initializations,   .) Evolution of the solutions during the iterations and
⋆) Final reconstructed objects.

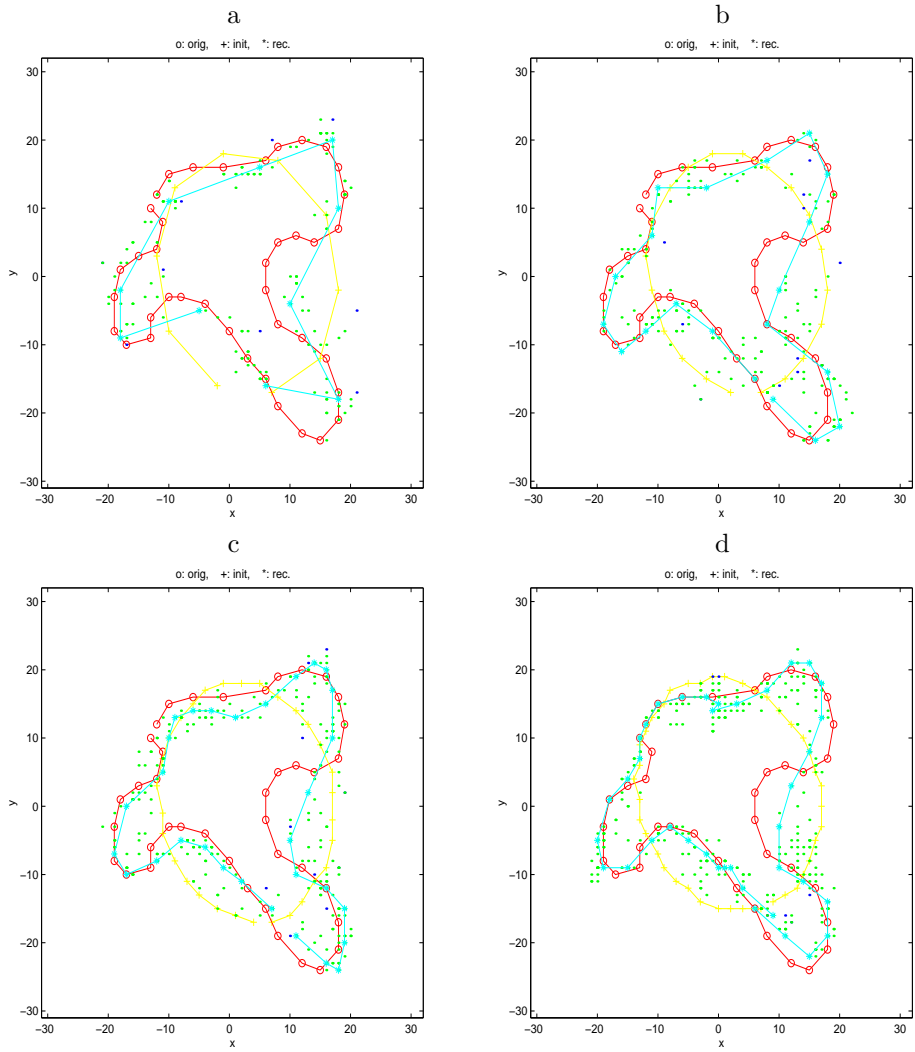

Figure 5: Reconstruction results with different number of vertices $N$ of the same object. The number of vertices of the original object is $N = 40$.
a) $N = 10$, b) $N = 20$ c) $N = 30$ and d) $N = 40$.

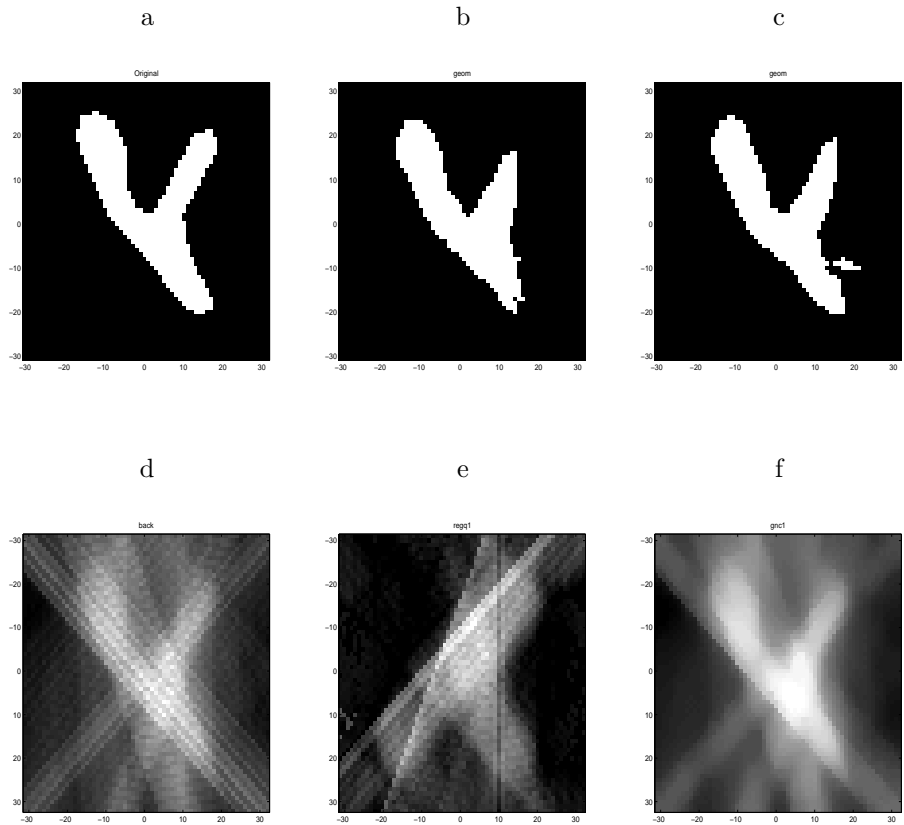

Figure 6: A comparison with backprojection and some other classical methods:
a) Original objects,
b) Results obtained by the proposed method using the SA optimization algorithm,
c) Results obtained by the proposed method using the ICM algorithm,
d) Backprojection,
e) Gaussian Markov modeling MAP estimation and
f) Compound Markov modeling and GNC optimization algorithm.

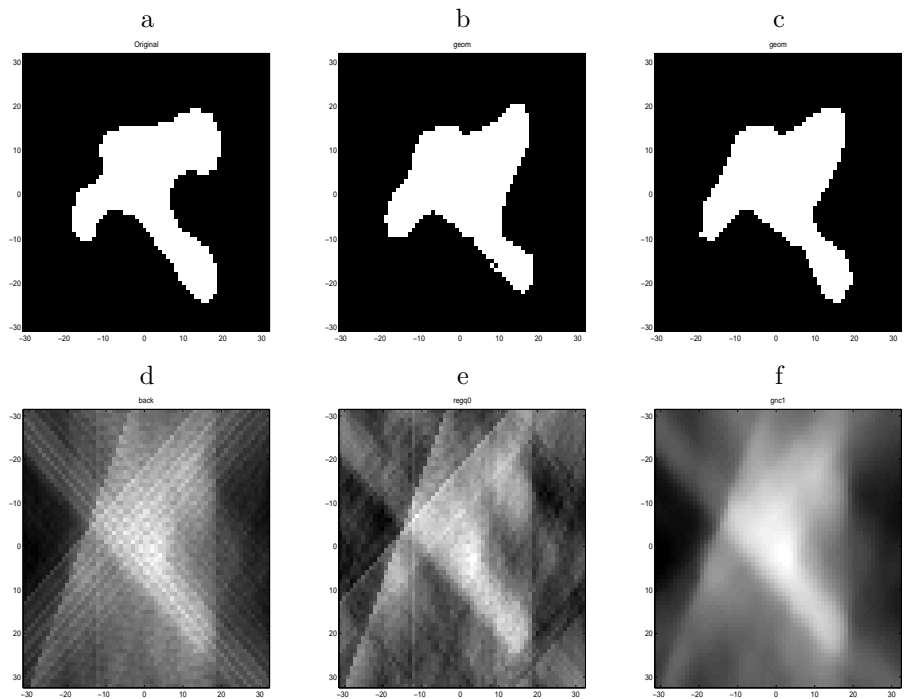

Figure 7: A comparison with backprojection and some other classical methods:
a) Original objects,
b) Results obtained by the proposed method using the SA optimization algorithm,
c) Results obtained by the proposed method using the ICM algorithm,
d) Backprojection,
e) Gaussian Markov modeling MAP estimation and
f) Compound Markov modeling and GNC optimization algorithm.

## 4. Conclusions

A new method for tomographic image reconstruction of a compact binary object from a small number of its projections is proposed. The basic idea of the proposed method is to model the compact binary object as a polygonal region whose vertices coordinates are estimated directly from the projections using the Bayesian MAP estimation framework or equivalently by optimizing a regularized criterion.

Unfortunately, this criterion is not unimodal. To find the optimized solution two algorithms have been proposed:
– a global optimization algorithm based on simulated annealing (SA) and
– a local descent-based method based on the Iterated Conditional Modes (ICM) algorithm proposed originally by Besag, with a good initialization obtained by using a moment based method.

The first algorithm seems to give entire satisfaction. The second can also give satisfaction, but it may also be plugged in a local minimum. In both algorithms the main cost calculation is due to the calculus of the variation of the criterion when one of the vertices coordinates is changed. We have written an efficient program to do this [29, 30].

An extension of this work to 3D image reconstruction with small number of conic projections is in preparation [31, 32]. The final objective of the proposed method is for non destructive testing (NDT) image reconstruction applications where we can use not only X-rays but also ultrasound or Eddy currents or a combination of them [33, 34, 11] to localize and to characterize more accurately any anomalies (air bulbs) in metallic structures.